\documentclass[prb,twocolumn]{revtex4-1} % Two column
\usepackage{amsmath}  % needed for \tfrac, \bmatrix, etc.
\usepackage{amsfonts} % needed for bold Greek, Fraktur, and blackboard bold
\usepackage{siunitx} %\SI[separate-uncertainty = true]{1.0(1)e-3}{\second}
\usepackage{verbatim}
\usepackage{color}
\usepackage[pdftex]{graphicx}
\usepackage{epstopdf} 
\usepackage{hyperref}

\begin{document}

% Be sure to use the \title, \author, \affiliation, and \abstract macros
% to format your title page.  Don't use lower-level macros to  manually
% adjust the fonts and centering.

\title{Non-radiative mid-range wireless power transfer: An experiment for senior physics undergraduates}
% In a long title you can use \\ to force a line break at a certain location.

\author{S. D. Delichte}
\author{Y. J. Lu}

\author{J. S. Bobowski}
\email{jake.bobowski@ubc.ca} % optional
%\altaffiliation[permanent address: ]{101 Main Street, 
%  Anytown, USA} % optional second address
% If there were a second author at the same address, we would put another 
% \author{} statement here.  Don't combine multiple authors in a single
% \author statement.
\affiliation{Department of Physics, University of British Columbia, Kelowna, British Columbia, Canada V1V 1V7}
% Please provide a full mailing address here.

% See the REVTeX documentation for more examples of author and affiliation lists.

\date{\today}

\begin{abstract}
A wireless power transfer experiment suitable for senior physics undergraduates that operates between \SI{3} and \SI{4}{\mega\hertz} is described and demonstrated in detail.  The apparatus consists of a pair of identical resonant coils that can be moved relative to one another.  A signal generator circuit is inductively coupled to the transmitting coil using a single loop of wire.  Likewise, a single loop of wire couples the receiving coil to a load impedance.  A matching circuit was used to tune the impedance of the system of coupled resonators to match the \SI[number-unit-product=\text{-}]{50}{\ohm} output impedance of the signal generator.  A low-cost vector network analyzer was used to characterize the system performance as the distance between the pair of coils was changed.  When the distance between the  transmitting and receiving coils was small, a double resonance emerged.  The frequency difference between the pair resonances was inversely proportional to the distance between the coils.  After the system was fully characterized, it was operated with up to \SI{20}{\watt} of incident power.  Our measurements revealed that, at weak coupling, the fraction of the incident power lost to radiation increases linearly with the distance between the coils.  Finally, we were able to transmit enough power to dimly light a \SI[number-unit-product=\text{-}]{60}{\watt} incandescent light bulb.
\end{abstract}
% AJP requires an abstract for all regular article submissions.
% Abstracts are optional for submissions to the "Notes and Discussions" section.

\maketitle % title page is now complete

\section{Introduction}

In 2007, a team of researchers at the Massachusetts Institute of Technology (MIT) led by Marin Solja\v{c}i\'{c} developed an efficient non-radiative wireless power transfer (WPT) system for mid-range distances.\cite{Soljacic:2007}  Here, mid-range implies that power is wirelessly transferred over a distance that is several times greater than the largest dimension of the transmitter/receiver device.  The scheme relies on the resonant coupling from a high-$Q$ receiver to a high-$Q$ transmitter.  This resonant coupling is acheived via overlap of the non-radiative envanescent fields of the pair of devices. To enhance the reach of these evanescent waves, the transmitter/receiver is designed to be many times smaller than the free-space resonant wavelength.\cite{Karalis:2008}  

Mid-range WPT can be implemented via a coupling to either evanescent electric or magnetic fields.  Magnetic coupling offers two key advantages: (1) Most common materials are nonmagnetic.  Therefore, objects placed between the transmitting and receiving resonators will typically only weakly affect the power transfer efficiency.\cite{Karalis:2008} (2) High-$Q$ resonators with evanescent magnetic fields that extend far outside the bodies of the resonant elements can be made very easily using inexpensive and easy-to-find materials.  This point is particularly relevant to the development of novel projects for undergraduate laboratories. In this work, identical resonant transmitter and receiver coils were made using \SI[number-unit-product=\text{-}]{1/4}{inch} copper tubing that can be found at most hardware stores.

\begin{figure*}
\centering{
(a)\quad\includegraphics[keepaspectratio, width=17 cm]{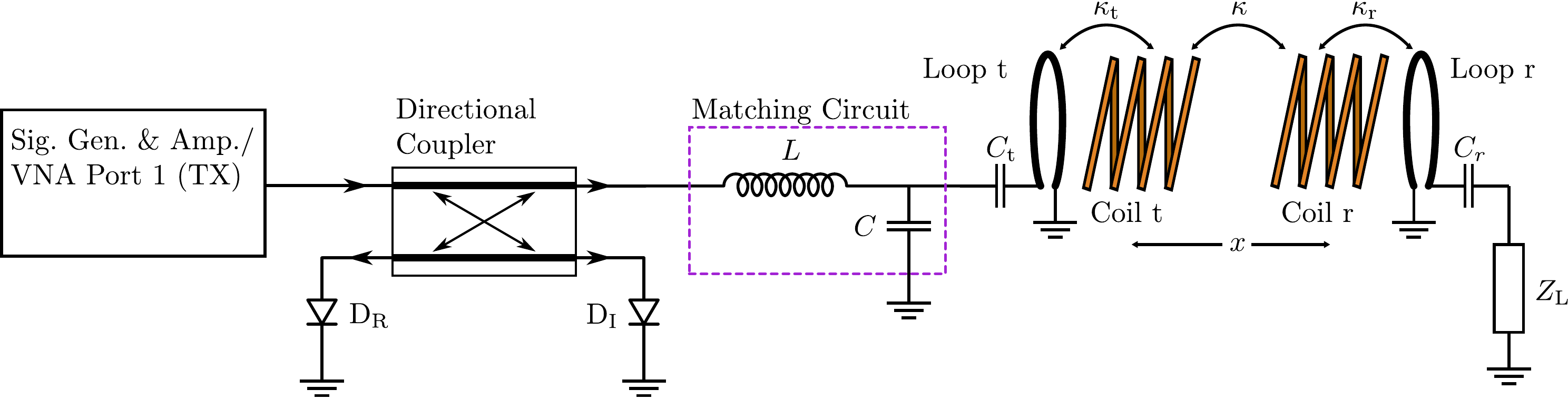}\vskip 7pt
(b)\quad\includegraphics[keepaspectratio, width=14 cm]{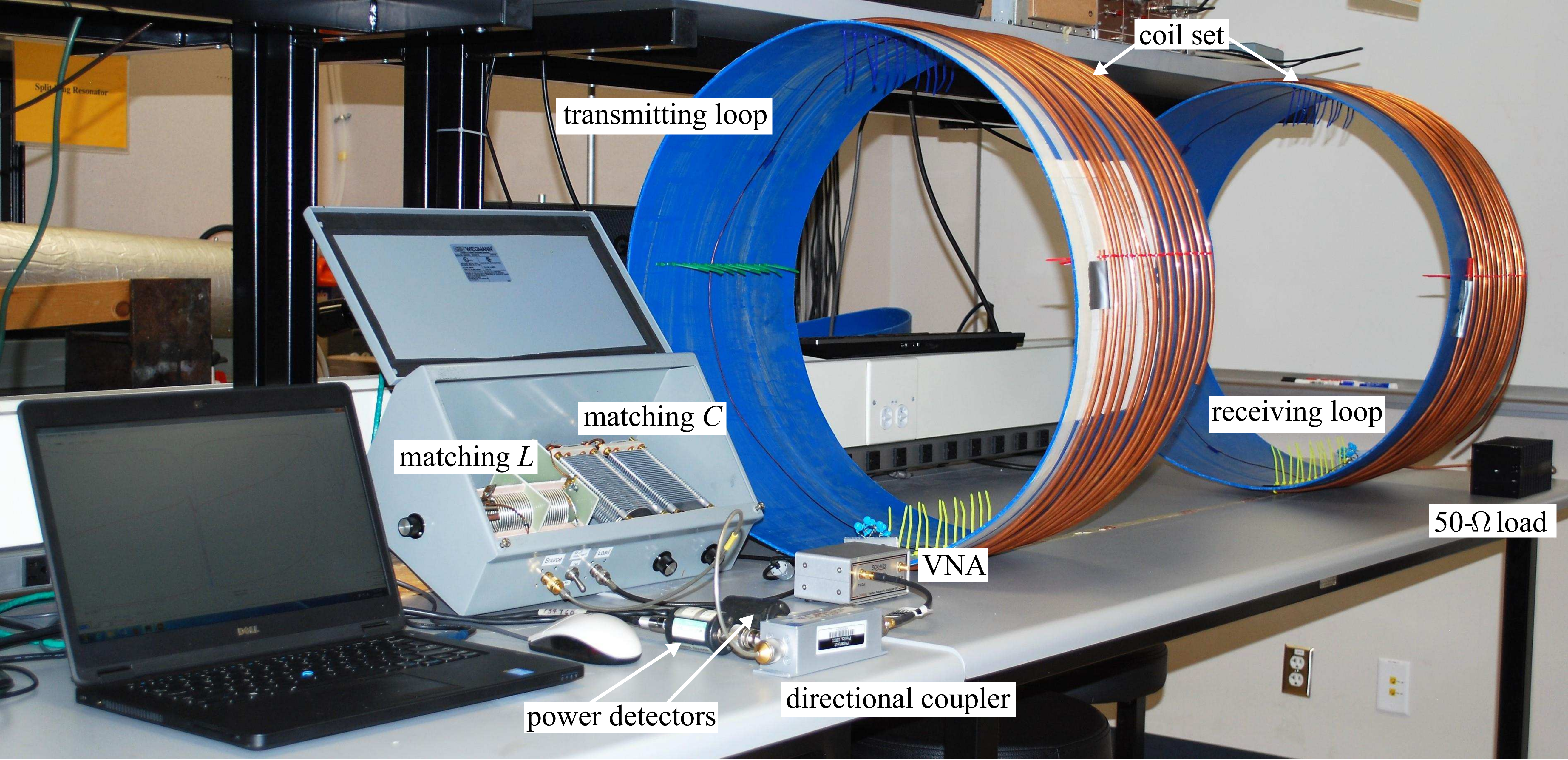}\vskip 3pt
(c)\quad\includegraphics[keepaspectratio, width=14 cm]{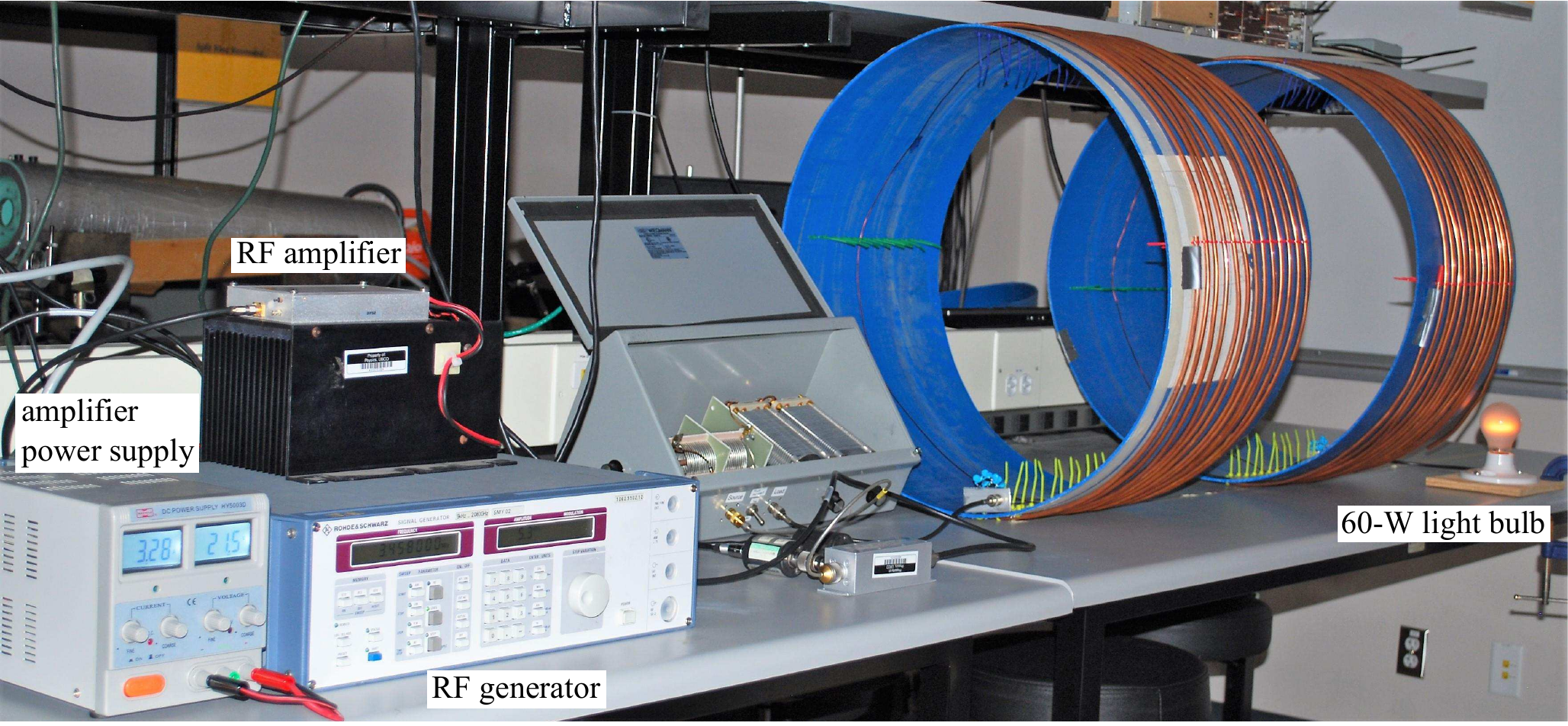}}
\caption{\label{fig:apparatus}(a) Schematic diagram of the WPT setup.  The signal source can be either Port 1 of a VNA or the amplified output of a signal generator.  The two diodes labeled $\mathrm{D_R}$ and $\mathrm{D_I}$ represent diode detectors used to measure the reflected and incident power, respectively.   As shown in (b), ``Loop t'' and ``Loop r'' are located within ``Coil t'' and ``Coil r'', respectively. (b) Digital photograph of the WPT setup while using the low-cost VNA and a \SI[number-unit-product=\text{-}]{50}{\ohm} load.  (c) A digital photograph of the WPT setup while using the signal generator, power amplifier, and a \SI[number-unit-product=\text{-}]{60}{\watt} incandescent light bulb.}
\end{figure*}

The WPT system used in the experiments presented in this work is shown schematically in Fig.~\ref{fig:apparatus}(a).  A single loop of wire, called ``Loop t'' in the figure, is connected to a signal generator via an impedance matching circuit.  The resonant frequency of the transmitting loop is tuned using capacitor $C_\mathrm{t}$.  Magnetic flux from this loop is inductively coupled to a nearby self-resonant coil referred to as ``Coil t''.  The mutual inductance between the transmitting loop and coil is characterized by coupling constant $\kappa_\mathrm{t}$.  The WPT receiver is made using an identical loop and coil with coupling constant $\kappa_\mathrm{r}$.  The receiver loop is terminated by a load impedance $Z_\mathrm{L}$.  The goal is to efficiently transfer power from the signal generator to $Z_\mathrm{L}$ while the transmitting and receiving coils are separated by a distance that can be several times greater than the diameter of the coils.  The coupling of the evanescent magnetic fields from one coil to the other is represented by the coupling constant $\kappa$ in Fig.~\ref{fig:apparatus}(a).

Circuit models that can be used to analyze (and optimize) WPT systems have been developed by several researchers.  See, for example, the review by Hui {\it et al.\/} and references therein.\cite{Hui:2014}  Although we do not repeat this type of analysis here, it is worth noting that students that have completed a course in electronics should be able to reproduce many of the key results.  In one particularly useful analysis, Cheon {\it et al.\/} examined the case in which the transmitting and receiving loops and coils all have the same resonant frequency.  In this case, the input impedance of the WPT system will equal to $Z_\mathrm{L}$, the impedance terminating the receiving loop, provided that the coupling constants satisfy the condition $\kappa_\mathrm{t}\kappa_\mathrm{r}/\kappa=1$.  If, under these conditions, $Z_\mathrm{L}$ is matched to the output impedance of the signal generator (typically \SI{50}{\ohm}), none of the incident power will be reflected back to the generator.\cite{Cheon:2011}

This impedance matching condition can be used to illustrate why the so-called four-coil system (transmitting loop and coil and receiving loop and coil) can be used for efficient mid-range WPT while a two-coil system, such as a simple transformer, is ineffective at these distances.  At mid-range distances, the coupling constant $\kappa$ in a two-coil system is inversely proportional to the cube of the distance $x$ between the coils.  Furthermore, the power transfer efficiency $\eta$ is proportional to $\kappa^2$ such that $\eta\propto x^{-6}$.  Therefore, as $x$ is increased to mid-range distances, the power transfer efficiency drops rapidly.\cite{Hui:2014}  In the four-coil system, however, the additional coupling constants $\kappa_\mathrm{t}$ and $\kappa_\mathrm{r}$ offer a mechanism to tune the power transfer efficiency even when a relatively large separation between the transmitting and receiving coils results in a small $\kappa$ value.  As long as the product $\kappa_\mathrm{t}\kappa_\mathrm{r}$ can similarly be reduced so as to satisfy the matching condition $\kappa_\mathrm{t}\kappa_\mathrm{r}/\kappa=1$, efficient power transfer can be achieved at mid-range distances.

In this paper, we describe our experiments with a simple four-coil WPT setup.  We used an inexpensive vector network analyzer (VNA) to characterize the system at low powers (\SI{-17}{dBm}, \SI{20}{\micro\watt}) and an RF power amplifier, bi-directional coupler, and power meters for measurements at high powers (up to \SI{20}{\watt}).  In our experiments, the coupling constants $\kappa_\mathrm{t}$ and $\kappa_\mathrm{r}$ were kept fixed and we achieved impedance matching using an external circuit as was done by Duong and Lee in Ref.~\onlinecite{Duong:2015}.  

We first measured how the magnitude of the signal reflected back to the generator, $\left\vert S_{11}\right\vert$, varied with the distance $x$ between the transmitting and receiving coils.  We then showed that, at all distances investigated, $\left\vert S_{11}\right\vert$ could be tuned to be zero at the resonant frequency using the impedance-matching circuit.  At small values of $x$, the system exhibited two distinct resonances.  We showed that the frequency splitting between the resonances was inversely proportional to $x$.  We also investigated the sensitivity of the system's matching conditions to variations in the load impedance $Z_\mathrm{L}$.  Finally, while operating at high powers, we measured the power-transfer efficiency as a function $x$ and used this information to deduce how much power was lost to radiation.  After fully characterizing the WPT system, it was used to dimly light a \SI[number-unit-product=\text{-}]{60}{\watt} incandescent light bulb.   

As a teaching tool, the WPT experiment offers two key benefits. First, the experiment introduces students to the VNA.   VNAs are now ubiquitous in RF and microwave research laboratories around the world.  However, due to their high cost, these instruments are not typically found in undergraduate laboratories.  Because the WPT systems operate at relatively low frequencies ($<\SI{10}{\mega\hertz}$), an inexpensive VNA can be used to characterize and optimize the system performance.  We purchased the \SI{1}{\kilo\hertz} to \SI{1.3}{\giga\hertz} \mbox{DG8SAQ VNWA 3} from SDR-Kits for \$600 USD and used it for all of the VNA measurements described here.\cite{SDR-Kits}  

The second key benefit is that this experiment appeals to a subset of physics undergraduates with a special interest in applied physics.  The goal of most of the projects offered to students in a physics lab course is to either measure a physical property of a material, measure a physical constant, or to demonstrate an interesting physical phenomenon.  In contrast, while still being rich in physics, the objective of the project described here is to optimize an apparatus so that it can perform a useful function -- the efficient wireless transfer of power from source to load over a mid-range distance.  Furthermore, the fact that mid-range WPT remains an extremely active area of research helps drive student interest in the project.    

\section{Design of the four-coil WPT system}
The design of the coils used in our experiments was inspired by the WPT apparatus developed by D.~Sherman and his students in the Advanced Physics Lab course at Cornell College in Mount Vernon, Iowa.\cite{Sherman:2011}     A list of the all of the equipment and materials used in our experiments is given in the appendix.  Where appropriate, possible vendors are suggested and estimates of costs given.

The transmitting and receiving coils were made from a \SI[number-unit-product=\text{-}]{7.58}{meter} length of standard \SI[number-unit-product=\text{-}]{1/4}{inch} copper tubing typically used for compressed air and water supply lines.  As shown in Fig.~\ref{fig:apparatus}, the copper tubing was wrapped around the midsection of \SI[number-unit-product=\text{-}]{55}{gallon} industrial plastic drum and held in place using plastic zip ties.  The finished coils consisted of 12 turns of copper tubing with a diameter of \SI{58}{\centi\meter} and a spacing of \SI{1.9}{\centi\meter} between each turn.

The transmitting and receiving loops were made from \SI[number-unit-product=\text{-}]{18}{AWG} copper magnet wire.  A single loop of wire was run along the inside wall of the plastic drums and held in place with zip ties and tape.  Series capacitors $C_\mathrm{t}$ and $C_\mathrm{r}$ were used to set the resonant frequency of the two loops.  Each of $C_\mathrm{t}$ and $C_\mathrm{r}$ were made using nine \SI{470}{\pico\farad}, \SI{6.3}{\kilo\volt} ceramic capacitors.  Three sets of three series capacitors were combined in parallel to make a net capacitance of \SI{470}{\pico\farad} capable of handling the power required to make the filament of a \SI[number-unit-product=\text{-}]{60}{\watt} incandescent light bulb glow. 

The tunable impedance-matching circuit was made using a \SIrange[range-phrase=--, range-units=single]{0}{22}{\micro\henry} variable air-roller inductor and a parallel combination of two \SIrange[range-phrase=--, range-units=single]{22}{1017}{\pico\farad} variable air-gap capacitors.  For the high-power measurements, the signal was supplied by a  Rohde \& Schwarz SMY02 signal generator and a \SI[number-unit-product=\text{-}]{30}{\watt} power amplifier manufactured by Mini-Circuits.  A bi-directional coupler was placed between the amplifier and matching circuit so that the incident and reflected power could be measured.  For all of the low-power measurements, the signal generator and amplifier were replaced by the VNA from SDR-Kits.  Since the VNA was only used to make reflection coefficient $\left(S_{11}\right)$ measurements, only \mbox{port 1} (TX) of the device was used in our experiments. 

\begin{figure*}
\centering{(a)\includegraphics[width=0.94\columnwidth]{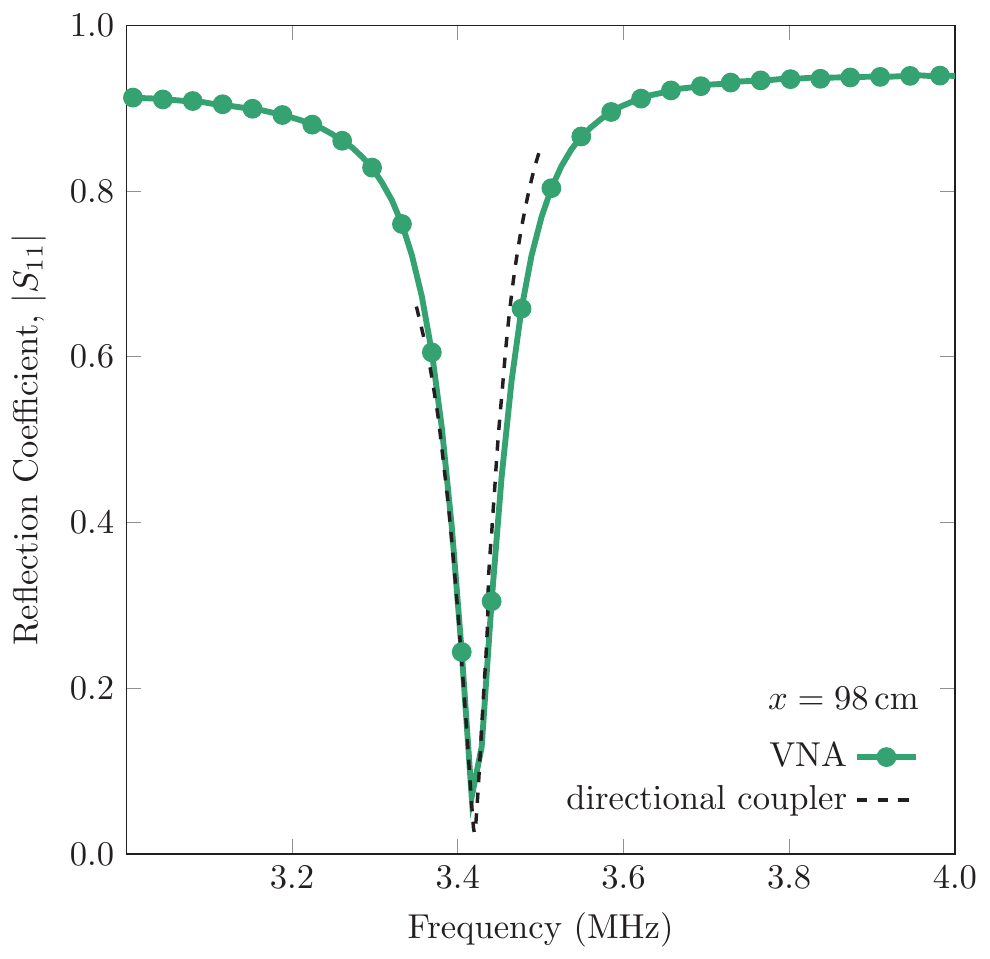}\qquad (b)\includegraphics[width=0.94\columnwidth]{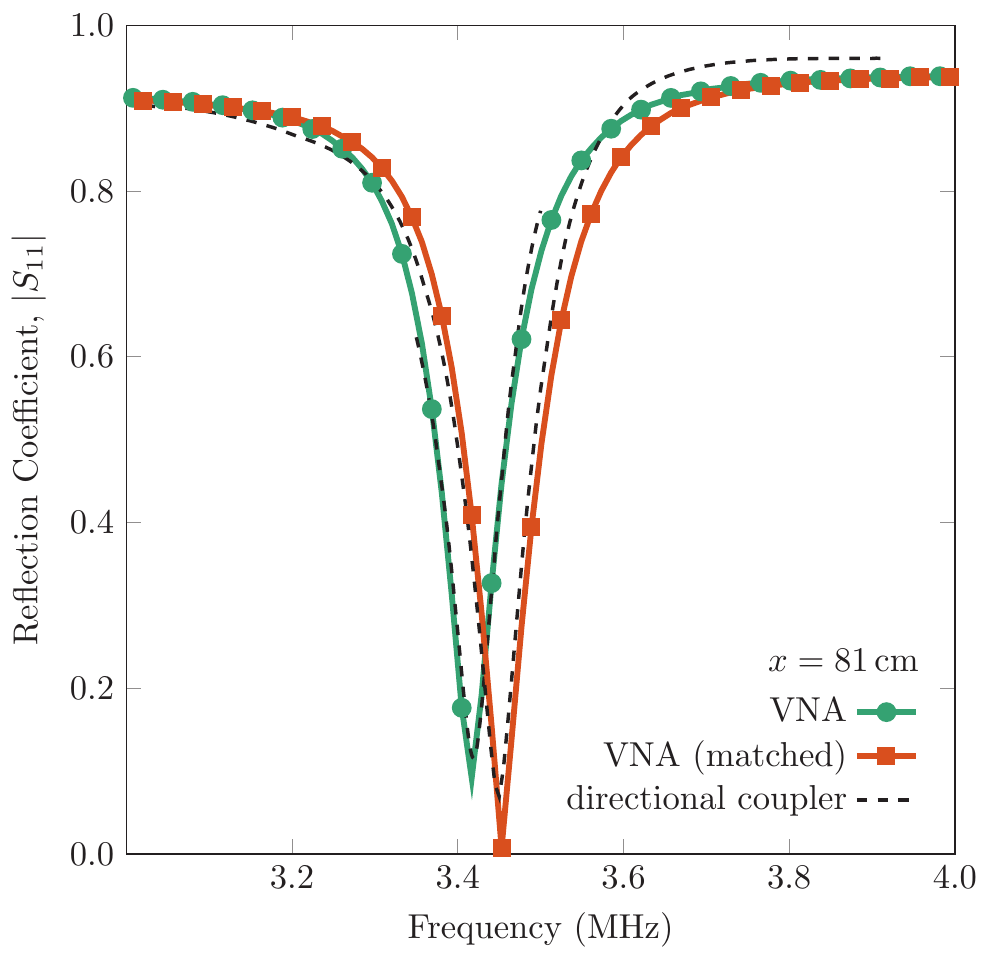}\\
(c)\includegraphics[width=0.94\columnwidth]{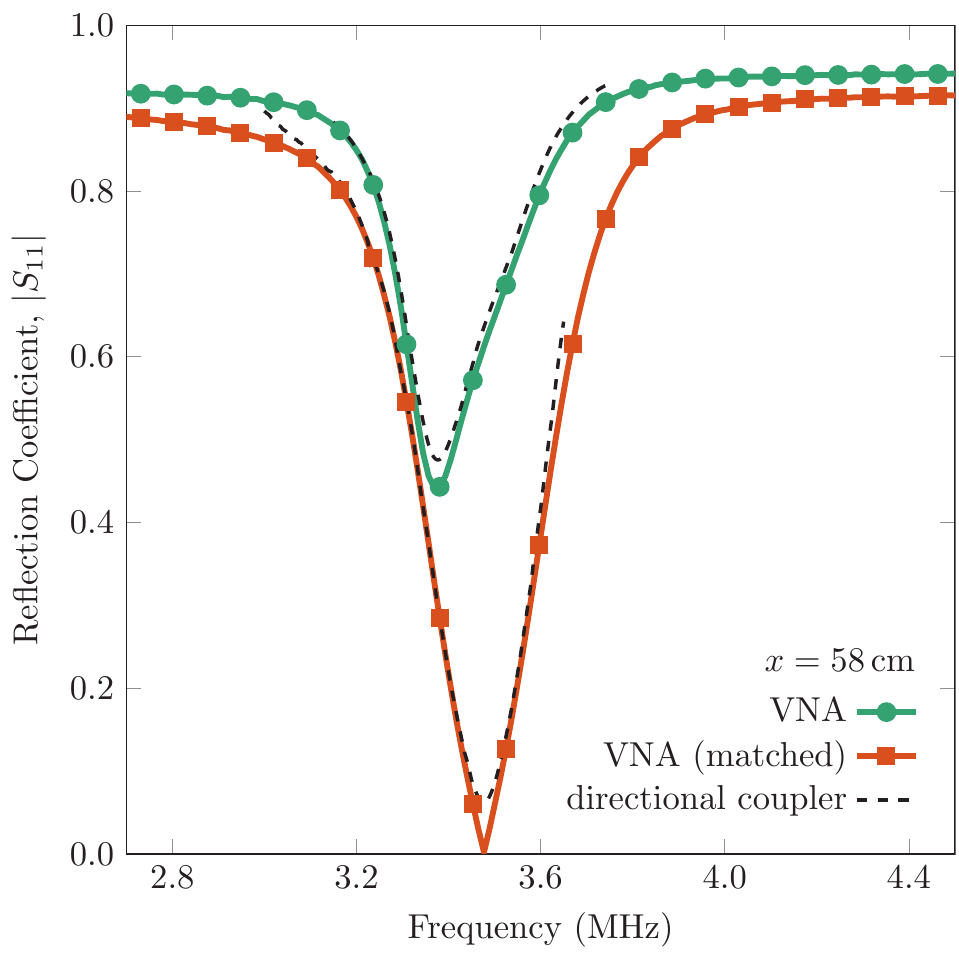}\qquad (d)\includegraphics[width=0.94\columnwidth]{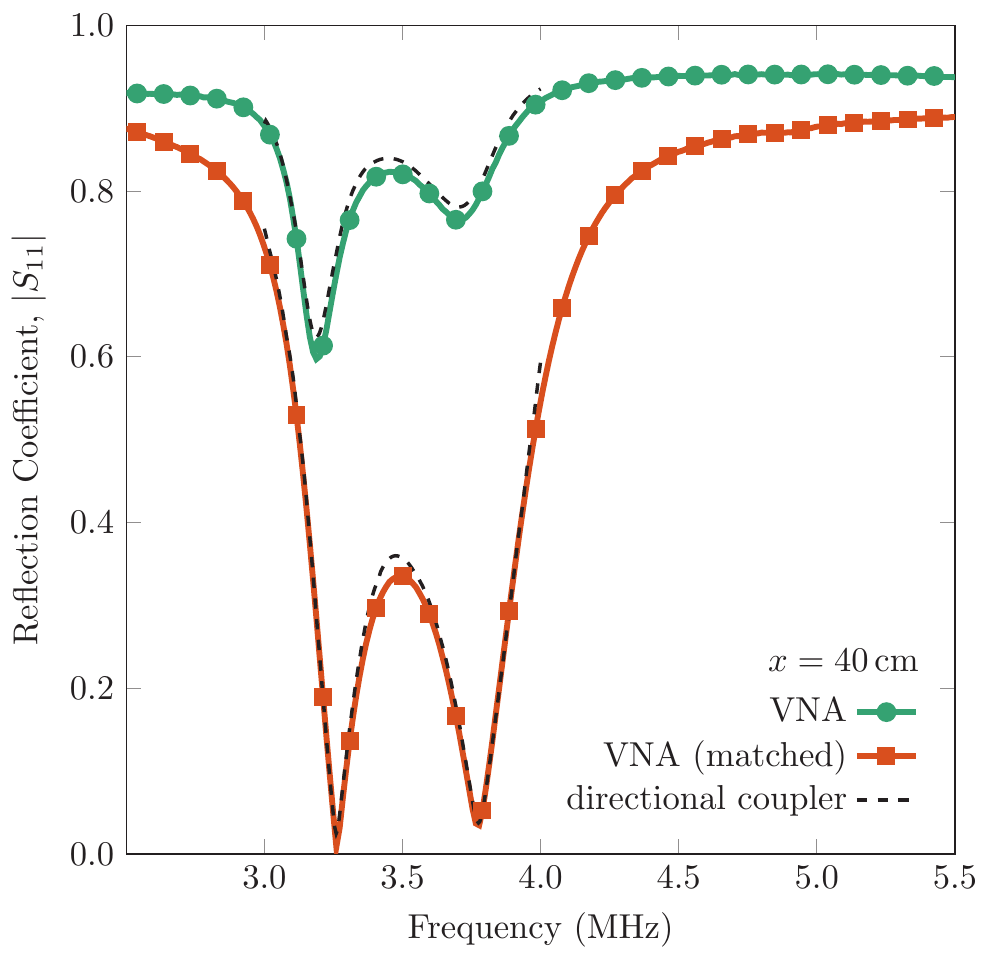}}
\caption{\label{fig:S11vsDist}  $\left\vert S_{11}\right\vert$ versus the distance $x$ between the centers of the two coils.  The green circles represent $\left\vert S_{11}\right\vert$ measured using the VNA when $L$ and $C$ of the matching circuit were set to achieve a good match at about \SI{3.4}{\mega\hertz} when $d=\SI{98}{\centi\meter}$.  The orange squares represent $\left\vert S_{11}\right\vert$ measured using the VNA when $L$ and $C$ of the matching circuit were readjusted to achieve a good match at each value of $x$.  For clarity, only a fraction of the all the data points measured for each curve are shown.  The dashed lines represent $\left\vert S_{11}\right\vert$ measured using the bi-directional coupler while operating the system at high powers ($\sim\SI{10}{\watt}$). (a) $x=\SI{98}{\centi\meter}$. (b) $x=\SI{81}{\centi\meter}$. (c) $x=\SI{58}{\centi\meter}$. (d) $x=\SI{40}{\centi\meter}$.}
\end{figure*}

\section{WPT system performance}
This section describes the measurements used to characterize the performance of the WPT system.

\subsection{Impedance matching and coil separation}\label{sec:matching}
If \mbox{$Z=R+jX$} is the effective impedance of the four-coil system terminated by $Z_\mathrm{L}$, then the impedance \mbox{$Z_\mathrm{in}=R_\mathrm{in}+jX_\mathrm{in}$} looking into the matching circuit is given by
\begin{align}
R_\mathrm{in}&=\frac{R}{\left(\dfrac{R}{\left\vert Z\right\vert}\right)^2+\left\{\omega\left\vert Z\right\vert C\left[1-\dfrac{X/\left\vert Z\right\vert}{\omega\left\vert Z\right\vert C}\right]\right\}^2}\label{eq:Rin}\\
X_\mathrm{in}&=\omega L-\frac{\omega\left\vert Z\right\vert^2C \left[1-\dfrac{X/\left\vert Z\right\vert}{\omega\left\vert Z\right\vert C}\right]}{\left(\dfrac{R}{\left\vert Z\right\vert}\right)^2+\left\{\omega\left\vert Z\right\vert C\left[1-\dfrac{X/\left\vert Z\right\vert}{\omega\left\vert Z\right\vert C}\right]\right\}^2},\label{eq:Xin}
\end{align}
where $\left\vert Z\right\vert=\sqrt{R^2+X^2}$.  Although Eqs.~(\ref{eq:Rin}) and (\ref{eq:Xin}) are somewhat complicated, the essential point is that $R_\mathrm{in}$ depends only on the value of $C$ while $X_\mathrm{in}$ is determined by both $C$ and $L$.  This fact is what makes the matching circuit, called an ``L-match'', so useful.  To get the maximum power transfer to the load impedance $Z_\mathrm{L}$, one tunes the matching circuit until none of the incident signal is reflected back to the source.  The reflection coefficient $S_{11}$ at the input of the matching circuit is given by
\begin{equation}
S_{11}=\frac{Z_\mathrm{in}-Z_0}{Z_\mathrm{in}+Z_0},\label{eq:S11}
\end{equation} 
where $Z_0=\SI{50}{\ohm}$ is the output impedance of the signal generator.  Therefore, the conditions required for zero reflection and, as a result, maximum power transfer are \mbox{$R_\mathrm{in}=Z_0$} and \mbox{$X_\mathrm{in}=0$}.

This impedance matching is achieved by first adjusting $C$ until $R_\mathrm{in}=Z_0$ and then tuning $L$ until $X_\mathrm{in}=0$.  The \mbox{DG8SAQ VNWA 3} data acquisition software allows the user to simultaneously display $R_\mathrm{in}$ and $X_\mathrm{in}$ as a function of frequency which makes the impedance matching process relatively straightforward.  For completeness, we note that a measurement of $S_{11}$ can also be used to determine $R_\mathrm{in}$ and $X_\mathrm{in}$.  Equation~(\ref{eq:S11}) can be used to show that
\begin{align}
\frac{R_\mathrm{in}}{Z_0}&=\frac{1-\left\vert S_{11}\right\vert^2}{1-2\Re\left[S_{11}\right]+\left\vert S_{11}\right\vert^2}\label{eq:RinZ0}\\
\frac{X_\mathrm{in}}{Z_0}&=\frac{2\Im\left[S_{11}\right]}{1-2\Re\left[S_{11}\right]+\left\vert S_{11}\right\vert^2},\label{eq:XinZ0}
\end{align}
where \mbox{$\left\vert S_{11}\right\vert^2=\Re\left[S_{11}\right]^2+\Im\left[S_{11}\right]^2$} and $\Re\left[S_{11}\right]$ and $\Im\left[S_{11}\right]$ are the real and imaginary parts of $S_{11}$, respectively.

Figure~\ref{fig:S11vsDist}(a) shows $\left\vert S_{11}\right\vert$ as a function of frequency as measured by the VNA after completing the impedance matching process.  For this measurement the center-to-center distance between the transmitting and receiving coils was $x=\SI{98}{\centi\meter}$ and the receiving loop was terminated with a load impedance $Z_\mathrm{L}=\SI{50}{\ohm}$.  At the system's resonant frequency of \SI{3.45}{\mega\hertz}, $\left\vert S_{11}\right\vert$ is nearly zero indicating that no power is being reflected back to the source (port 1 of the VNA).  Away from the resonant frequency, $\left\vert S_{11}\right\vert > 0.9$ indicating that almost all of the incident signal in reflected back to the source.  

For all VNA measurements, a calibration plane was established at the free end of a coaxial cable connected to port 1 of the VNA.  The calibration was done using the calibration standards (open, short, and \SI[number-unit-product=\text{-}]{50}{\ohm} load) supplied by SDR-Kits and the \mbox{DG8SAQ VNWA 3} data acquisition software.  The purpose of the calibration is to correct for any imperfections (losses and spurious reflections) in the cable and connectors used to connect the device under test to port 1 of the VNA.\cite{DaSilva:1978, Bobowski:2012}  

As shown in Fig.~\ref{fig:apparatus}(b), the VNA measurement was done with the bi-directional coupler placed between port 1 of the VNA and the input of the matching circuit.  This was done in anticipation of the high-power measurements in which the signal generator and power amplifier were used in place of the VNA.  The dashed line shown in Fig.~\ref{fig:S11vsDist}(a) shows $\left\vert S_{11}\right\vert$ extracted while using an incident power of approximately \SI{10}{\watt}.  We automated this measurement using a LabVIEW program that stepped the signal generator through a range of frequencies and recorded the incident and reflected powers from the bi-directional coupler at each step.  It is important to note that the reflection coefficient is defined as a ratio of voltages, therefore, in terms of power \mbox{$\left\vert S_{11}\right\vert=\sqrt{P_\mathrm{r}/P_\mathrm{i}}$}, where $P_\mathrm{i}$ and $P_\mathrm{r}$ are the root-mean-square (rms) incident and reflected powers, respectively.  In Fig.~\ref{fig:S11vsDist}(a), the $\left\vert S_{11}\right\vert$ curves measured using the VNA and the directional coupler are in good agreement.  

\begin{figure}[t]
\centering{\includegraphics[width=0.94\columnwidth]{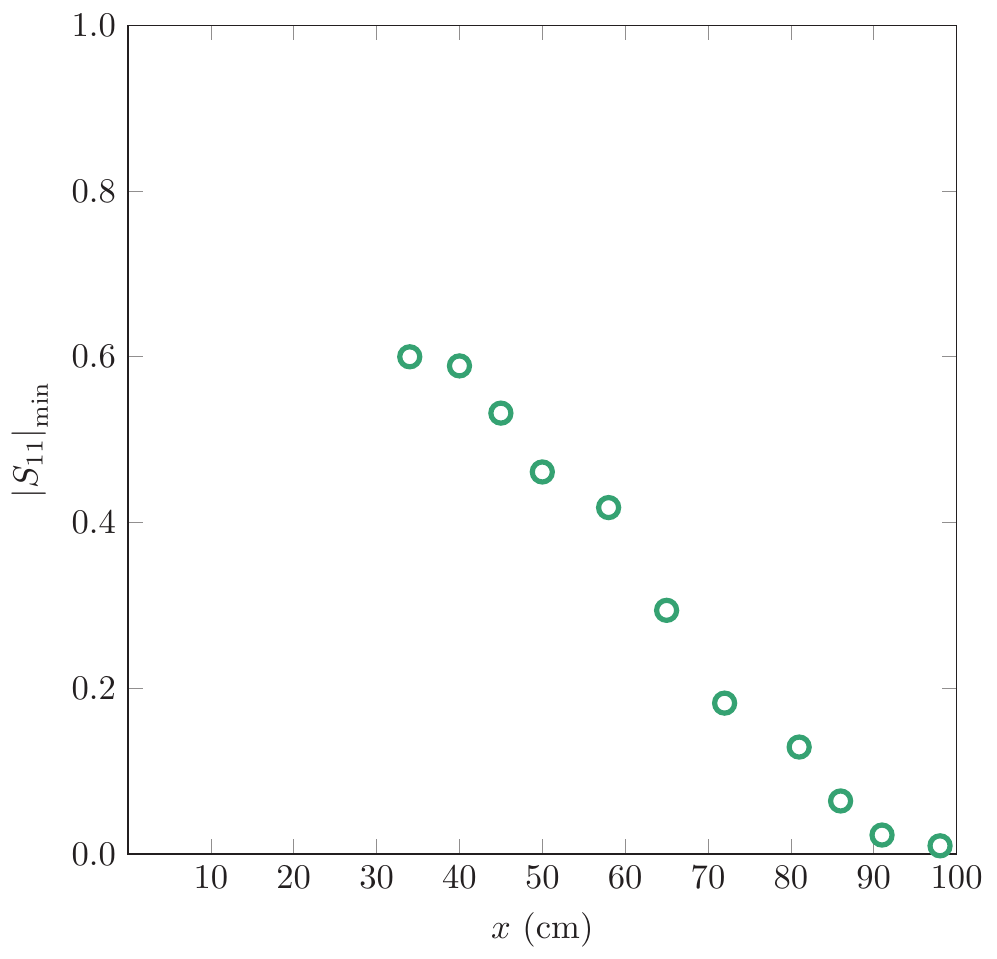}}
\caption{\label{fig:S11min}Minimum of $\left\vert S_{11}\right\vert$ as a function of the separation distance $x$ between the two coils.  The matching circuit was tuned to produce a good match to \SI{50}{\ohm} when $x=\SI{98}{\centi\meter}$.  See the data shown by green circles in Fig.~\ref{fig:S11vsDist}.}
\end{figure}

Next, without making any changes to $L$ and $C$ of the matching circuit, the distance $x$ between the transmitting and receiving coils was reduced to \SI{81}{\centi\meter} and the low- and high-power $\left\vert S_{11}\right\vert$ measurements were repeated.  The results are shown by the green circles and dashed line in Fig.~\ref{fig:S11vsDist}(b).  The minimum value of $\left\vert S_{11}\right\vert$ at the resonant frequency is slightly increased indicating that more of the incident power is reflected back to the signal source.  Clearly, decreasing the distance between the transmitting and receiving coils increases the coupling constant $\kappa$.  However, since the transmitting and receiving loops are fixed within their respective coils, the coupling constants $\kappa_\mathrm{t}$ and $\kappa_\mathrm{r}$ remain unchanged.  Recall that the matching condition for a four-coil WPT system is given by \mbox{$\kappa_\mathrm{t}\kappa_\mathrm{r}/\kappa=1$}.  Therefore, changing $\kappa$ while keeping the other coupling constants fixed is expected to degrade the impedance match of the WPT system.  As the green circles in Figs.~\ref{fig:S11vsDist}(c) and (d) show, the mismatch becomes even greater as the distance between the coils is further reduced.  Notice that, at the closest spacing shown in Fig.~\ref{fig:S11vsDist}(d), a double resonance emerges.  The origin and properties of this double resonance will be discussed in Sec.~\ref{sec:double}.  Figure~\ref{fig:S11min} shows that, if a good match is established at $x=\SI{98}{\centi\meter}$, then the minimum of $\left\vert S_{11}\right\vert$ increases approximately linearly as $x$ is decreased.

\begin{figure*}
\centering{(a)\includegraphics[width=0.94\columnwidth]{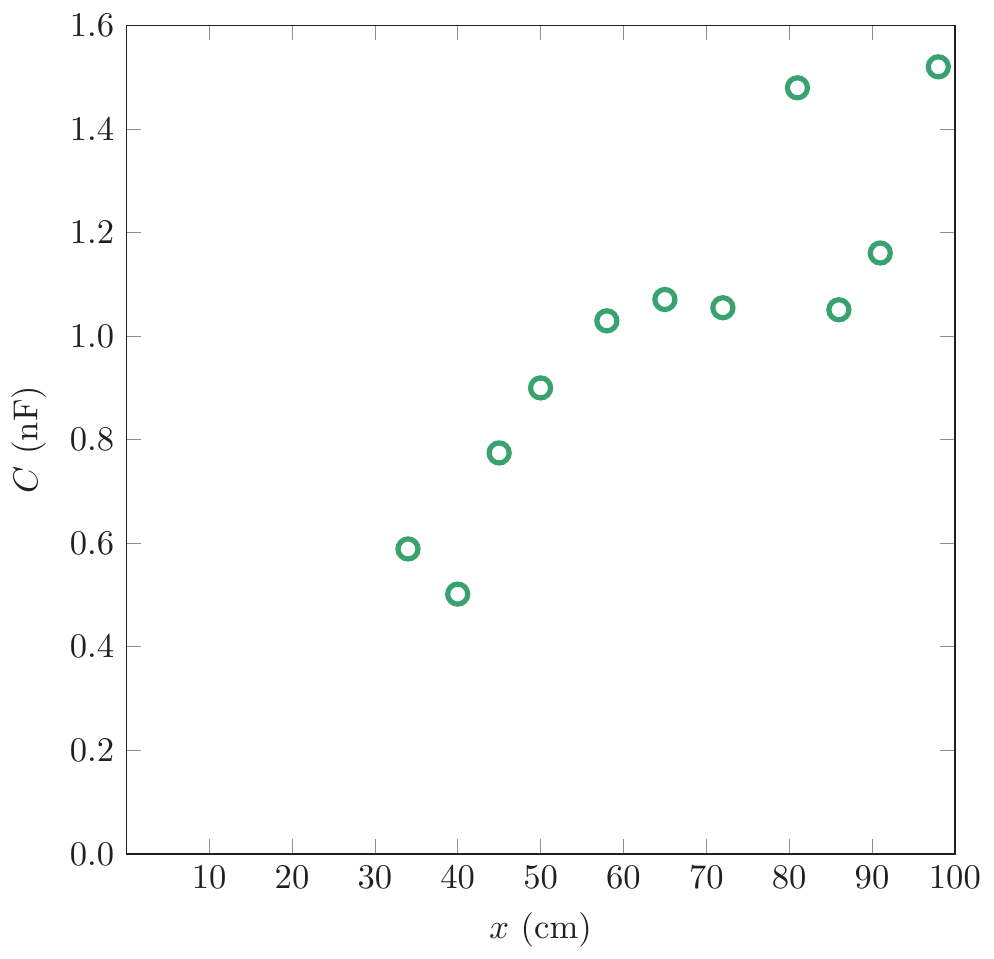}\qquad (b)\includegraphics[width=0.94\columnwidth]{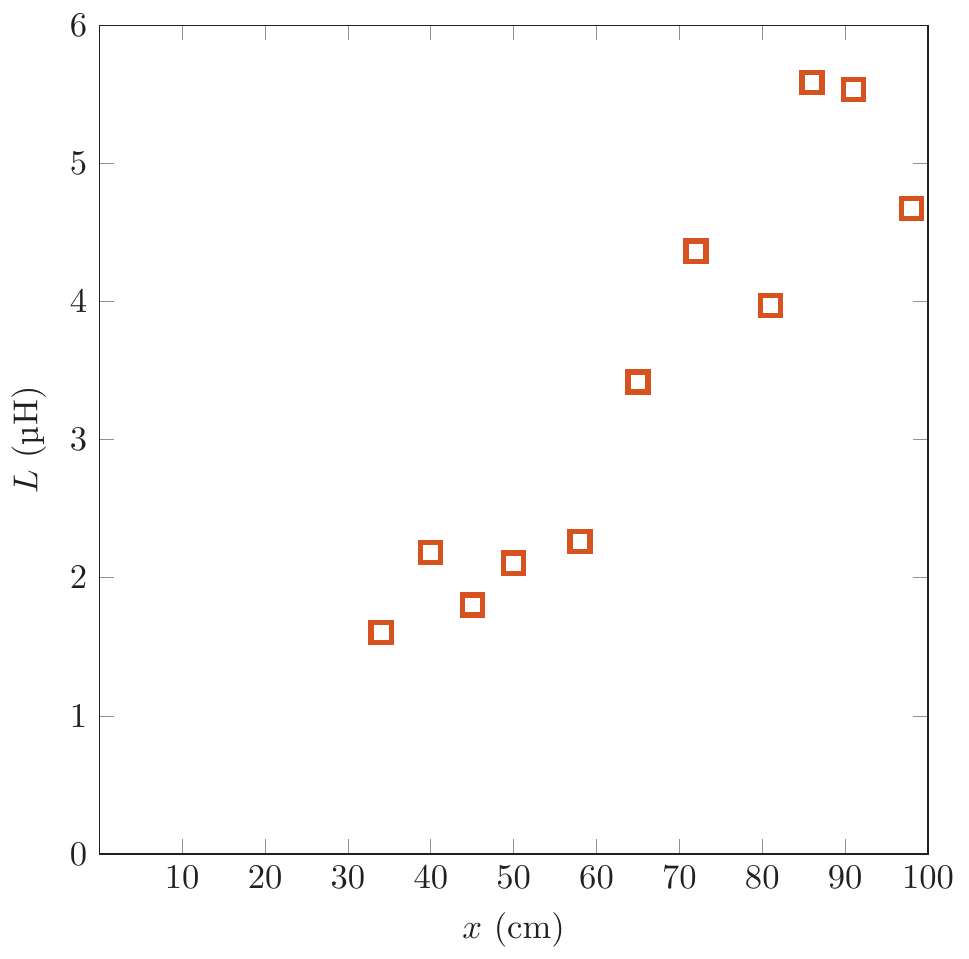}}
\caption{\label{fig:CandL}The values of the matching (a) capacitance $C$ and (b) inductance $L$ as a function of the separation distance $x$ between the resonant transmitting and receiving coils.}
\end{figure*}

We conclude this section by showing that the matching circuit can be used to reestablish a condition of zero reflection at all of the values of $x$ examined.  For $x=\SI{81}{}$, \SI{58}{}, and \SI{40}{\centi\meter} the orange squares in Fig.~\ref{fig:S11vsDist} show the measured $\left\vert S_{11}\right\vert$ after tuning the matching circuit.  At all three positions, $\left\vert S_{11}\right\vert\approx 0$ at the resonant frequency.  Notice also that the resonant frequency changes slightly as $x$ is varied.  In Fig.~\ref{fig:CandL}, we show the values of $C$ and $L$ required to achieve a match to the source impedance of \SI{50}{\ohm} for values of $x$ ranging from \SI{34}{} to \SI{98}{\centi\meter}. 

The measurements of $C$ and $L$ were made by using the VNA to measure the complex reflection coefficient of a signal incident on the capacitor/inductor.  For an impedance that is completely imaginary, \mbox{$\left\vert S_{11}\right\vert=1$} and Eq.~(\ref{eq:XinZ0}) simplifies to
\begin{equation}
\frac{X_\mathrm{in}}{Z_0}=\frac{\Im\left[S_{11}\right]}{1-\Re\left[S_{11}\right]}.\label{eq:Xin1}
\end{equation}
One can then determine the capacitance from \mbox{$C=1/\left(\omega X_\mathrm{in}\right)$} or the inductance from \mbox{$L=X_\mathrm{in}/\omega$}.  Using this method, the values of $C$ and $L$ required to achieve impedance matching were found to increase approximately linearly with $x$.  Notice also that both plots in Fig.~\ref{fig:CandL} have $y$-intercepts that are close to zero.  This result is consistent with the condition that the four-coil WPT system should be matched to the source impedance when \mbox{$\kappa_\mathrm{t}\kappa_\mathrm{r}/\kappa=1$}.  In our experimental geometry $\kappa_\mathrm{t}$ and $\kappa_\mathrm{r}$ are both fixed and expected to be close to one.  In this case, the WPT system is expected to be impedance matched, without the use of the matching circuit, when $\kappa\approx 1$.  Under these circumstances, all of the magnetic field lines that pass through the transmitting coil would also have to pass through the receiving coil, a condition that would necessarily be satisfied if $x$ could be set to zero.

\subsection{Double resonance when $x$ is small}\label{sec:double}

\begin{figure*}
\centering{(a)\includegraphics[width=0.94\columnwidth]{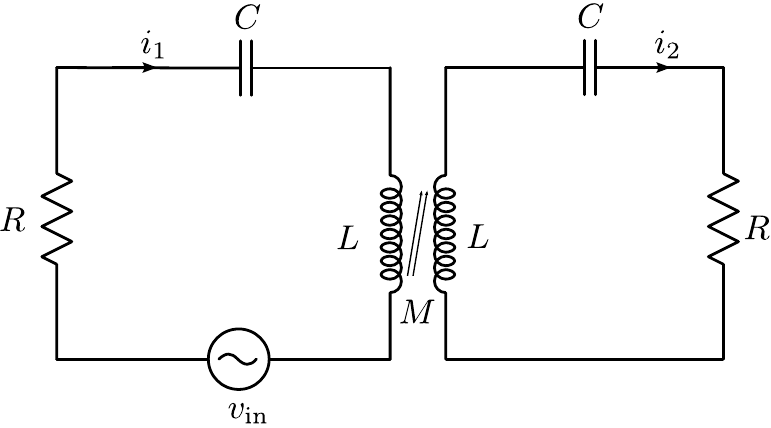}\qquad (b)\includegraphics[width=0.94\columnwidth]{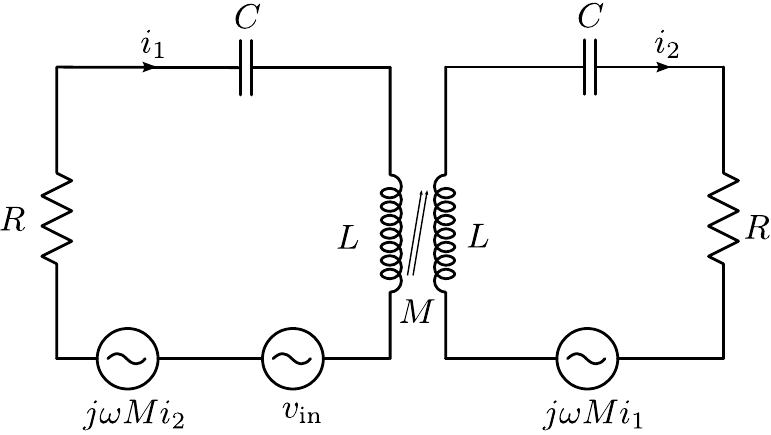}}
\caption{\label{fig:toyModel}(a) A pair of identical resonators, modeled as $LRC$-circuits, coupled via shared magnetic flux.  (b) Equivalent circuit for the coupled resonators.  The left-hand circuit induces an emf proportional to $i_1$ in the right-hand circuit.  The right-hand circuit induces an emf proportional to $i_2$ in the left-hand circuit.}
\end{figure*}
\begin{figure*}
\centering{(a)\includegraphics[width=0.94\columnwidth]{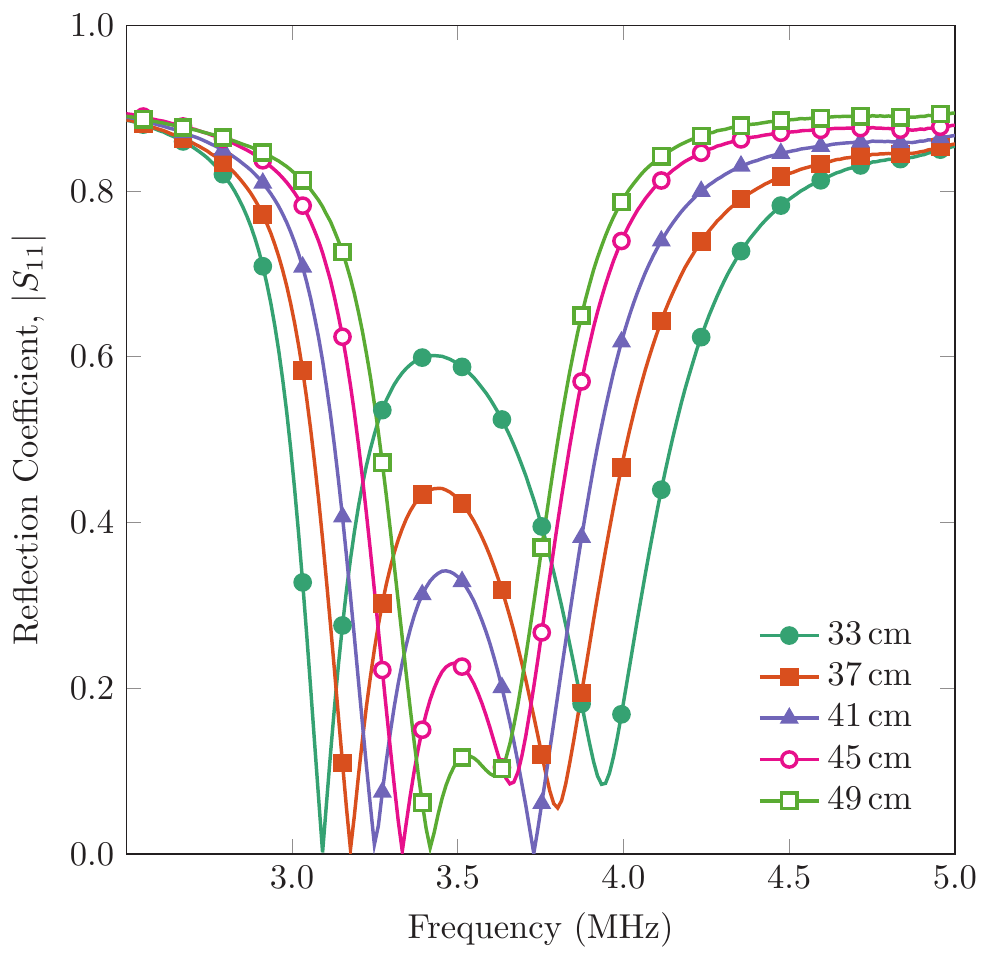}\qquad (b)\includegraphics[width=0.94\columnwidth]{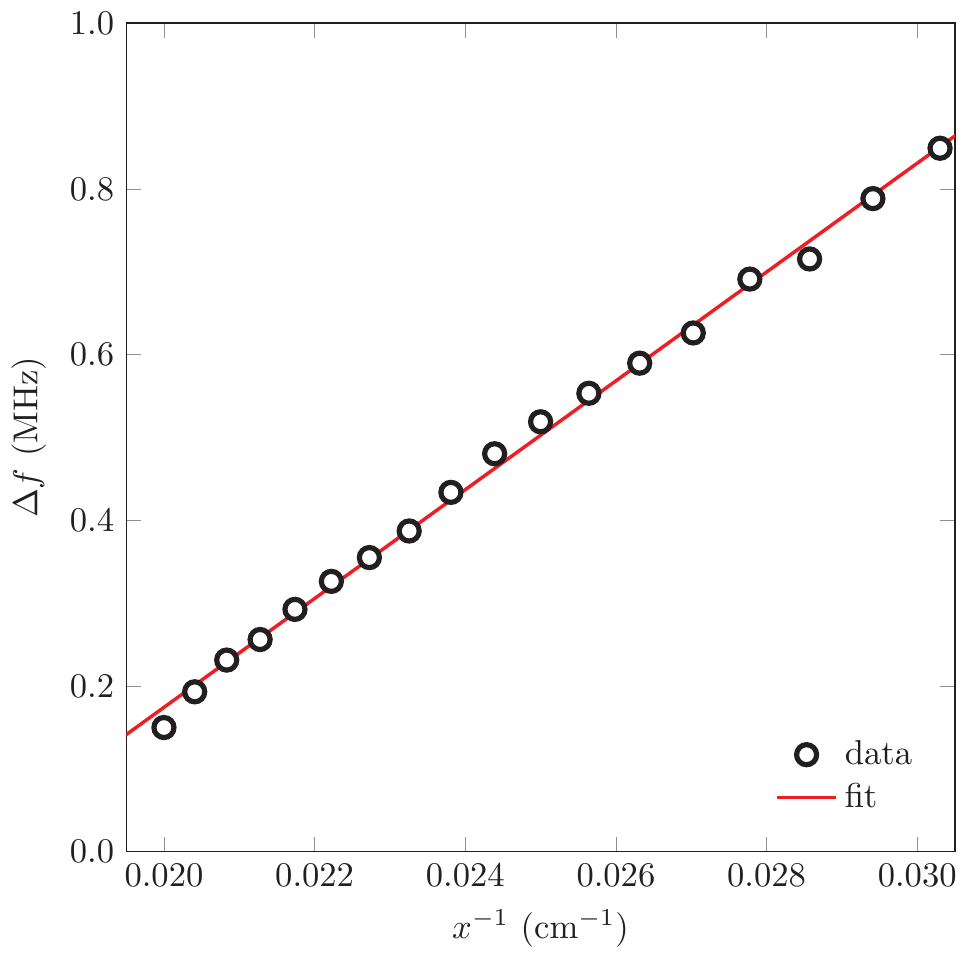}}
\caption{\label{fig:strongCoupling}(a) The $\left\vert S_{11}\right\vert$ double resonance measured at five values of $x$ between \SI{33}{} and \SI{49}{\centi\meter}.  At each value of $x$, the matching capacitance and inductance were tuned to produce a good match at the frequency of the low-frequency minimum.  (b) A plot of the difference in frequency $\Delta f$ between the minima of the double resonance versus the inverse of the distance between the coils.}
\end{figure*}

We next turn our attention to the double resonance that was observed in Fig.~\ref{fig:S11vsDist}(d) when $x=\SI{40}{\centi\meter}$.  When identical resonators are coupled in some manner, the coupled system exhibits a double resonance and the frequency splitting depends on the strength of the coupling.\cite{Mehdizadeh:1983, Liu:2011, Dubreuil:2017}  Figure~\ref{fig:toyModel}(a) shows a simple circuit model in which identical resonators, treated as $LRC$ circuits, are coupled via shared magnetic flux.  In this model, it is assumed that the left circuit is driven by a voltage source supplying $v_\mathrm{in}$.  The mutual inductance $M$ between the two circuits is given by $\kappa L$ where $\kappa$ is a coupling constant.  By Faraday's law of induction, an emf proportional to $i_1$ is induced in the right circuit and an emf proportional to $i_2$ is induced in the left circuit.  Figure~\ref{fig:toyModel}(b) shows an equivalent circuit for the coupled resonators that explicitly includes in the induced voltages.

Kirchhoff voltage loops around each of the circuits yields the following system of equations
\begin{align}
v_\mathrm{in}+j\omega M i_2-i_1\left(R+\frac{1}{j\omega C}+j\omega L\right)&=0\\
j\omega M i_1-i_2\left(R+\frac{1}{j\omega C}+j\omega L\right)&=0,
\end{align}
which can be solved for the two unknown currents $i_1$ and $i_2$.  Completing this analysis shows that the currents each exhibit a double resonance with resonant frequencies given by
\begin{equation}
\omega^\prime_\pm\approx \omega_0\left(1\pm\frac{\kappa}{2}\right),
\end{equation}
where we have assumed weak coupling and underdamped (high $Q$) resonators.\cite{Dubreuil:2017}  In this expression, \mbox{$\omega_0=1/\sqrt{LC}$} is the resonant frequency of the individual resonators in the zero-coupling limit.  The frequency splitting, therefore, is proportional to the coupling strength and given by
\begin{equation}
\Delta\omega=\omega^\prime_+ - \omega^\prime_-\approx\kappa\omega_0.
\end{equation}

The transmitting and receiving coils of the WPT system were designed to be identical and their coupling is determined by the spacing of the coils.  Once the coils are close enough that $\kappa\omega_0$ becomes appreciable, a double resonance should emerge.  Figure~\ref{fig:strongCoupling}(a) shows $\left\vert S_{11}\right\vert$ measured at five different spacings ranging from $x=\SI{33}{}$ to \SI{49}{\centi\meter}.  At each value of $x$, the impedance matching circuit was tuned to make $\left\vert S_{11}\right\vert=0$ at the lower of the two resonant frequencies.  The double resonance is clearly observed in each measurement and, as expected, the greatest frequency splitting occurs at the smallest $x$ (i.e., at the strongest coupling).  This measurement was done at a total of 18 different values of $x$ and Fig.~\ref{fig:strongCoupling}(b) shows that the frequency splitting varies linearly with $x^{-1}$.  These data suggest that, at small separation distances, the coupling strength $\kappa$ is proportional to $x^{-1}$. 

\subsection{The dependence of $\left\vert S_{11}\right\vert$ on $Z_\mathrm{L}$}

\begin{figure*}
\centering{(a)\includegraphics[width=0.94\columnwidth]{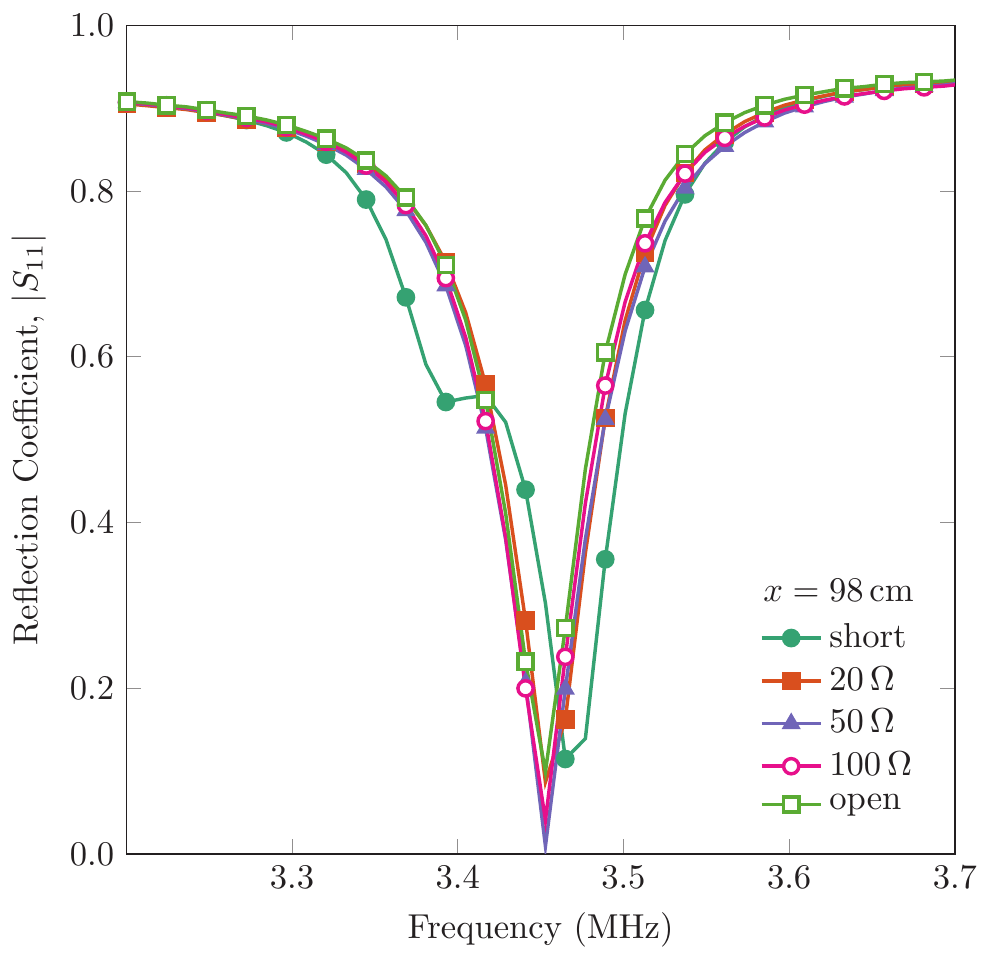}\qquad (b)\includegraphics[width=0.94\columnwidth]{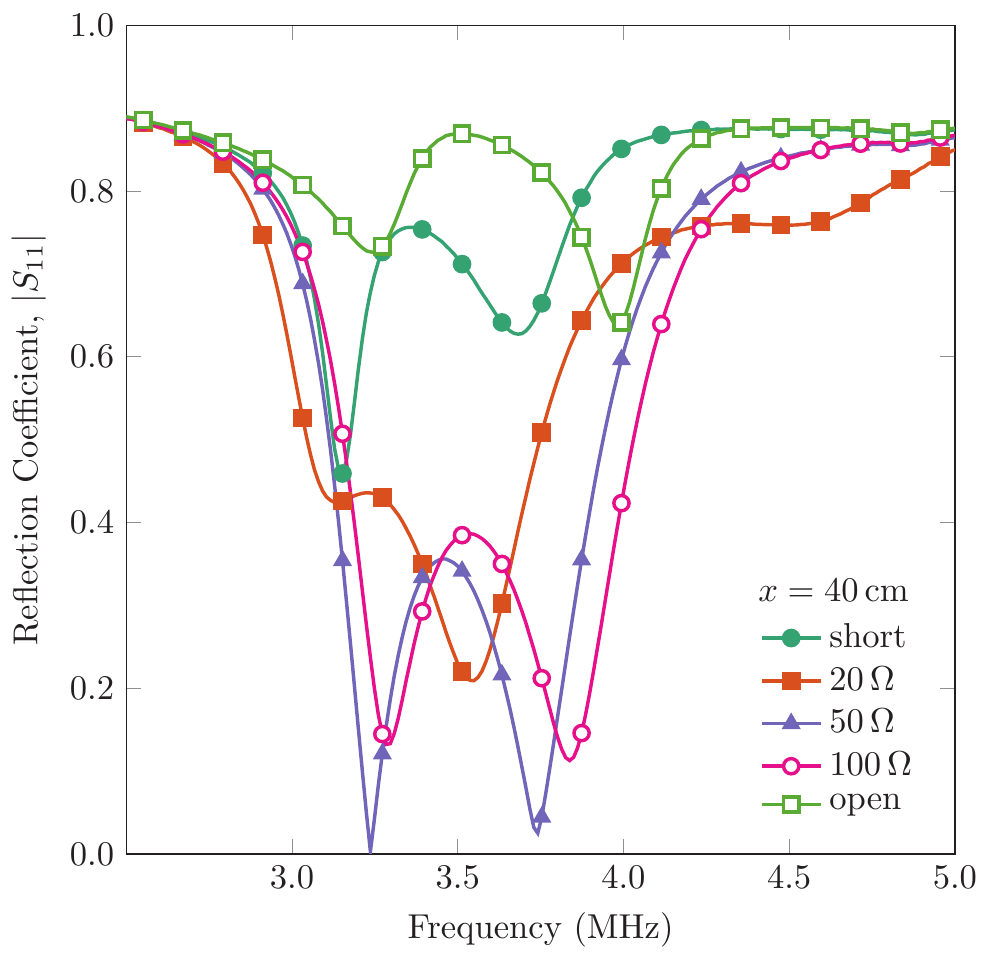}}
\caption{\label{fig:loadDependence}The dependence of $\left\vert S_{11}\right\vert$ on the load impedance $Z_\mathrm{L}$ terminating the receiving loop (Loop r in Fig.~\ref{fig:apparatus}(a)) when (a) $x=\SI{98}{\centi\meter}$ and (b) $x=\SI{40}{\centi\meter}$.}
\end{figure*}

In this section, we consider how the load impedance $Z_\mathrm{L}$ terminating the receiving loop affects $\left\vert S_{11}\right\vert$.  In a real-world application, one might imagine a fixed transmitter delivering power to a mobile receiver.  Furthermore, depending on the particular situation, the receiver may be required to drive a variable load impedance.  Intuitively, one expects a variable load to present more of a challenge when the WPT coils are relatively strongly coupled and be less of an issue when the coupling is weak.

We measured the dependence of $\left\vert S_{11}\right\vert$ on $Z_\mathrm{L}$ when \mbox{$x=\SI{98}{\centi\meter}$} and then again when \mbox{$x=\SI{40}{\centi\meter}$}.  For these measurements, we started with \mbox{$Z_\mathrm{L}=\SI{50}{\ohm}$} and tuned the matching circuit such that \mbox{$\left\vert S_{11}\right\vert=0$} at the resonant frequency.  We then varied $Z_\mathrm{L}$ and, without making any changes to the matching circuit, observed the response of $\left\vert S_{11}\right\vert$.  We considered only resistive loads (\SI{20}{}, \SI{50}{} and \SI{100}{\ohm}) and open- and short-circuits.

Figure~\ref{fig:loadDependence}(a) shows that, when \mbox{$x=\SI{98}{\centi\meter}$} (weak coupling), the changes to $\left\vert S_{11}\right\vert$ are modest and a relatively good match to the source impedance, at the resonant frequency, is maintained for any value of $Z_\mathrm{L}$.  These measurements were repeated at \mbox{$x=\SI{40}{\centi\meter}$} and the results are shown in Fig.~\ref{fig:loadDependence}(b).  Due to the close spacing of the coils, the double resonance discussed in the previous section is seen in the data.  In stark contrast to the \SI{98}{\centi\meter} data, $\left\vert S_{11}\right\vert$ is seen to depend very strongly on the value of $Z_\mathrm{L}$ and a poor match to the source impedance results when $Z_\mathrm{L}$ deviates from \SI{50}{\ohm}.    

It is worth noting that this mid-range WPT system is designed for distances that are up to several times greater than the coil diameter.  With a coil diameter of \SI{58}{\centi\meter}, a separation distance of \SI{40}{\centi\meter} would not be typical in a practical application.  Therefore, it is reasonable to infer that variations of the load impedance $Z_\mathrm{L}$ would not typically be expected to degrade the performance of a commercial mid-range WPT system. 

\subsection{Power transfer efficiency}

Section~\ref{sec:matching} demonstrated that the bi-directional coupler can be used to make reliable measurements of incident and reflected power.  This section, describes our use of the directional coupler to determine the efficiency with which incident power was transferred to the load impedance $Z_\mathrm{L}$ as the coil separation distance $x$ was varied.  

At each value of $x$, we first used the VNA to tune the matching circuit such that \mbox{$\left\vert S_{11}\right\vert=0$} at the resonant frequency.  Next, the VNA was removed and replaced with the signal generator and power amplifier.  The frequency of the signal generator was set to the known resonant frequency and its output power, as measured by the power meter monitoring the ``forward'' port of the bi-directional coupler, was in the range from \SI{7.1}{} to \SI{7.4}{\watt}.  At this point, the signal generator frequency and the values of $C$ and $L$ in the matching circuit were fine tuned while monitoring the reflected power at the ``reflected'' port of the bi-directional coupler.  The fine tuning was done so as to achieve the lowest possible reflected power.  After fine tuning, the reflected power was typically \mbox{$\SI{4}{\milli\watt}\pm 50\% $}, which corresponds to a reflection coefficient of \mbox{$\left\vert S_{11}\right\vert\approx\SI{0.02}{}\pm \SI{0.01}{}$}.

\begin{figure}[t]
\centering{\includegraphics[width=0.94\columnwidth]{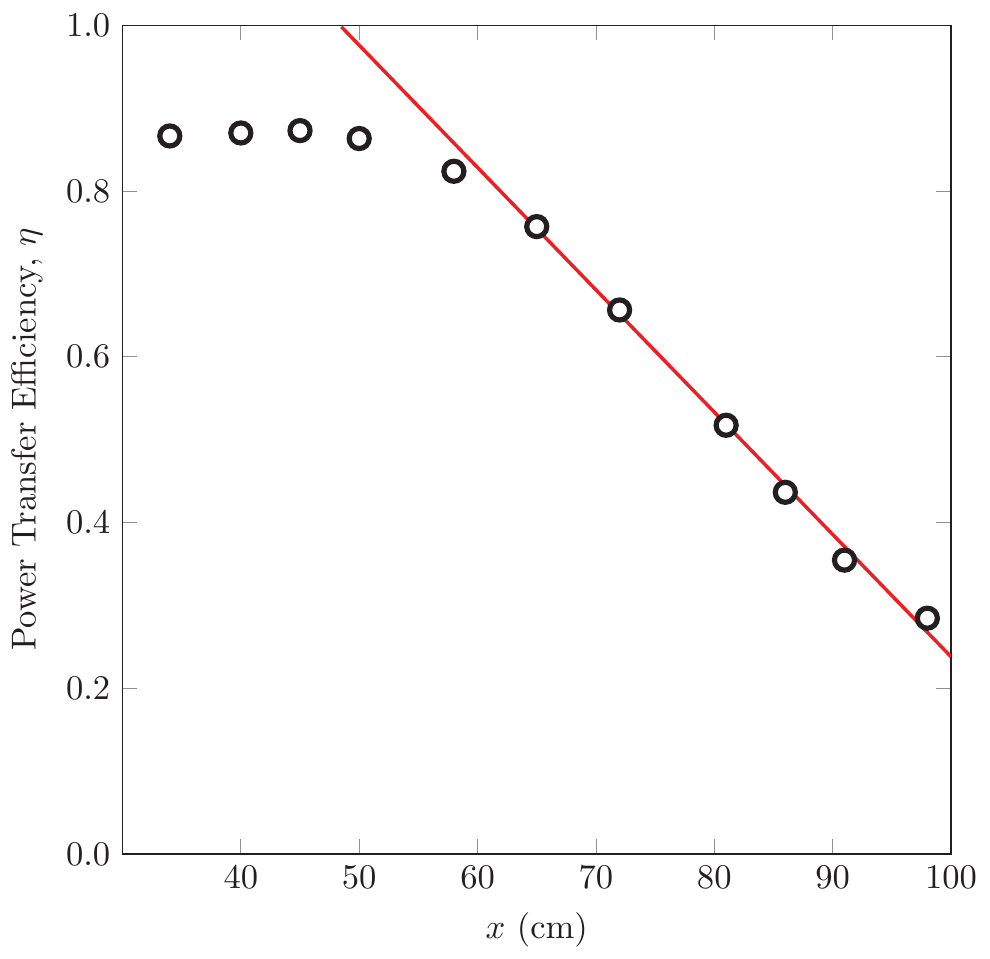}}
\caption{\label{fig:eta}The fraction of the incident power transmitted to the \SI[number-unit-product=\text{-}]{50}{\ohm} load impedance as a function of the coil separation distance $x$.}
\end{figure}

The impedance $Z_\mathrm{L}$ terminating the receiving loop was a \SI[number-unit-product=\text{-}]{50}{\ohm} load capable of dissipating \SI{100}{\watt}.  The voltage across $Z_\mathrm{L}$ was monitored using a digital oscilloscope and probes with a 10:1 attenuation setting.  The transmitted power $P_\mathrm{t}$ was then calculated from the measured rms voltage: \mbox{$P_\mathrm{t}=V_\mathrm{rms}^2/Z_\mathrm{L}$}.  Finally, the fraction of the incident power $P_\mathrm{i}$ that was deliver to $Z_\mathrm{L}$ was determined from \mbox{$\eta=P_\mathrm{t}/P_\mathrm{i}$}.  Figure~\ref{fig:eta} shows $\eta$ measured as a function of the coil separation $x$.

For separations less than \SIrange[range-phrase=--, range-units=single]{50}{60}{\centi\meter}, the power transfer efficiency was approximately constant at \mbox{$\SI{0.87}{}\pm\SI{0.02}{}$}.  For values of $x$ above \SI{60}{\centi\meter}, the efficiency decreased linearly with $x$ falling to \SI{0.28}{} at \mbox{$x=\SI{98}{\centi\meter}$}.  The fraction of reflected power, given by $\left\vert S_{11}\right\vert^2$, is negligible and losses in the copper coils and loops are likewise expected to be relatively small.  Therefore, the fraction of incident power that is not absorbed by $Z_\mathrm{L}$ must be lost predominantly to radiation.  That is, a fraction of the magnetic flux generated by the transmitting coil, rather than being coupled to the receiving coil, is radiated to free space.  Our measurements suggest that this radiated power increases linearly with coil separation beyond some critical value of $x$.

Finally, we replaced the \SI[number-unit-product=\text{-}]{50}{\ohm} termination with a \SI[number-unit-product=\text{-}]{60}{\watt} incandescent light bulb and attempted to transfer enough power to visibly light the bulb.  A coil separation \mbox{$x=\SI{58}{\centi\meter}$} was used and the incident power was \SI{20.4}{\watt}.  Before applying the high power, the VNA was used to tune the matching circuit while the light bulb was cold and the DC resistance of the filament was measured to be \SI{16.7}{\ohm}.  After applying the high power, the filament heats and its resistance increases enough that an improvement in performance can be gained by fine tuning the signal generator frequency  -- an implementation of the so-called adaptive frequency tuning.\cite{Sample:2011}  After adjusting the frequency to \SI{3.48}{\mega\hertz}, the rms voltage across the filament of the light bulb was measured to be \SI{42.4}{\volt} and the reflected power was \SI{5.4}{\watt}.  This reflected power could have been reduced by fine tuning $C$ and $L$ of the matching circuit, but this step was not completed.  Figure~\ref{fig:apparatus}(c) shows a photograph of the WPT apparatus being used to light the incandescent bulb.

To find the transmitted power, it was first necessary to determine the resistance of the heated filament.  We applied the same rms voltage to the light bulb at \SI{60}{\hertz} and measured the resulting current.  The filament resistance was found to be \SI{141}{\ohm}.  One might worry that, due to the skin effect, the resistance of the filament at \SI{3.48}{\mega\hertz} would be higher than it is at \SI{60}{\hertz} for the same temperature.  However, at \SI{2000}{\celsius} tungsten has a resistivity of \SI{67}{\micro\ohm\centi\meter} which gives a skin depth of \mbox{$\delta\approx\SI{220}{\micro\meter}$} at \SI{3.48}{\mega\hertz}.\cite{Jones:1926}  A typical coiled-coil tungsten filament in a \SI[number-unit-product=\text{-}]{60}{\watt} bulb has a diameter of \SI{46}{\micro\meter} which is much less than our calculated $\delta$ and suggests that the resistance measurement at \SI{60}{\hertz} should be a good approximation to the resistance at \SI{3.48}{\mega\hertz}.\cite{Agrawal:2011}  Therefore, the power transferred to the bulb was estimated to be \SI{12.8}{\watt}.  As a result, the fraction of power lost to radiation and dissipation within the WPT apparatus is estimated to be $\approx 0.11$ which is in reasonable agreement with the efficiency versus $x$ data shown in Fig.~\ref{fig:eta}. 

\section{Summary}\label{sec:summary}

A four-coil WPT apparatus was designed and constructed.  A low-cost vector network analyzer was used to characterize and optimize the system's performance.  Using a tunable impedance-matching circuit, the magnitude of the reflection coefficient $\left\vert S_{11}\right\vert$ could be set close to zero for a wide range of separation distances $x$ between the transmitting and receiving coils.  At small separation distances, a double resonance emerged due to the relatively strong coupling between the two identical resonant coils.  The frequency spacing between the two resonances increased linearly with the $x^{-1}$ suggesting that, at small separations, the coupling constant $\kappa$ is proportional to $x^{-1}$.  It was demonstrated that, at large values of $x$, the load impedance $Z_\mathrm{L}$ terminating the receiving loop has very little affect on $\left\vert S_{11}\right\vert$, whereas the affect is large when $x$ is small.  It is worth noting that, for WPT systems designed for mid-range distances, the double resonance and the strong dependence of $\left\vert S_{11}\right\vert$ on $Z_\mathrm{L}$ at small values of $x$ are not likely to be of great importance.  

Next, the VNA was replaced with a signal generator and power amplifier for high-power operation.  The efficiency with which incident power was transferred to $Z_\mathrm{L}$ was measured as a function of $x$.  The efficiency was constant at $\approx 0.87$ up to \mbox{$x=\SI{50}{\centi\meter}$} and then decreased linearly with $x$, falling to \SI{0.28}{} when \mbox{$x=\SI{98}{\centi\meter}$}.  Finally, the WPT system was used to illuminate a \SI[number-unit-product=\text{-}]{60}{\watt} light bulb.  A power of \SI{12.8}{\watt} was successfully transferred to the bulb over a coil separation distance of \SI{58}{\centi\meter}.

This experiment is particularly well suited for students with an interest in applied physics.  In contrast to most of the experiments offered to physics undergraduates, the goal of investigating mid-range WPT methods is fundamentally to develop real-world applications.  The experiment presented, however, is still rich in physics.

To conclude, we suggest some further investigations that could supplement or extend the measurements describe above.  (i) The coupling constants $\kappa_\mathrm{t}$ and $\kappa_\mathrm{r}$ could be varied by allowing the transmitting/receiving loops to be moved relative the their respective coil.  This modification would allow one to try to achieve the required impedance matching condition \mbox{$\kappa_\mathrm{t}\kappa_\mathrm{r}/\kappa=1$} without the need of a separate impedance matching circuit.  (ii) In our experiments, we varied only the separation distance of coils. One could also examine the system's response if the axes of the cylinders are displaced laterally or rotated with respect to one another.\cite{Duong:2015} (iii) For students with an interest in programming, automated adaptive tuning algorithms could be developed.  These could include matching circuits with switchable values of $C$ and $L$,\cite{Duong:2015} or an implementation of adaptive frequency tuning in which one takes advantage of the fact that the resonant frequency depends weakly on the coupling constants.\cite{Sample:2011}  (iv) As was done by D.~Sherman and his students, rather than purchase a relatively expensive RF power amplifier, one could design their own low-cost switch-mode amplifier for the high-power experiments.\cite{Sherman:2011}  (v) In our experiments, the load impedance $Z_\mathrm{L}$ (\SI[number-unit-product=\text{-}]{50}{\ohm} load or \SI[number-unit-product=\text{-}]{60}{\watt} bulb) was driven by the RF signal at approximately \SI{3.5}{\mega\hertz}.  In a real-world application, one would first rectify the signal and drive the final load with a DC voltage.  Students with an interest in RF electronics could investigate the design of a rectifier circuit suitable for a WPT system.\cite{Collado:2017}  (vi) Finally, it is also possible to design WPT systems that capacitively couple a transmitter and receiver via electric fields.\cite{Dai:2015}

\appendix*

\section{Parts and Suppliers}

This appendix provides a list of the equipment required to reproduce all parts of the WPT experiment described in this paper.  Where appropriate, vendors and cost estimates are also provided.  The VNA and matching circuit components are required for the low-power characterization and tuning of the WPT apparatus.  All other RF test equipment is required only for the high-power measurements.

{\it VNA} --  Research quality VNAs are generally very expensive and not found in undergraduate labs.  We used a USB VNA sold by SDR-Kits \mbox{(\url{https://www.sdr-kits.net/})}.  The \mbox{DG8SAQ VNWA 3} has a dynamic range of \SI{90}{\decibel} up to \SI{500}{\mega\hertz} and can be used up to \SI{1.3}{\giga\hertz} at a reduced dynamic range.  The data acquisition software for the \mbox{DG8SAQ VNWA 3} is freely available from the SDR-Kits website.  Its cost, including calibration kit, ranges from \$530 to \$640 depending on the options selected.

{\it Matching Circuit} -- To construct the impedance-matching circuit, two M73 \SIrange[range-phrase=--, range-units=single]{22}{1017}{\pico\farad} variable air-gap capacitors from Oren Elliott Products were combined in parallel \mbox{(\url{http://www.orenelliottproducts.com/})}.  At the time of purchase, these cost \$106 each.  The impedance-matching circuit also used a \SI{22}{\micro\farad} air-roller variable inductor from MFJ Enterprises which cost \$54 \mbox{(\url{http://www.mfjenterprises.com/})}.

{\it Signal Generator} -- Our measurements at high power made use of a Rohde \& Schwarz SMY02 \SI{9}{\kilo\hertz} to \SI{2080}{\mega\hertz} signal generator capable of outputting a maximum of \SI{19}{dBm}  (\SI{80}{\milli\watt}) of power.  A suitable substitute would be the B\&K Precision 4017A \SI{0.1}{\hertz} to \SI{10}{\mega\hertz} signal generator with an output power up to \SI{24}{dBm} (\SI{250}{\milli\watt}) \mbox{(\url{http://www.bkprecision.com/})}.  The manufacturer's suggested retail price is \$459.

{\it RF Power Amplifier} -- We used the Mini-Circuits \mbox{LZY-22+} \SI{100}{\kilo\hertz} to \SI{200}{\mega\hertz}, \SI{30}{\watt} amplifier \mbox{(\url{https://www.minicircuits.com/})}.  This amplifier can withstand both open- and short-circuits at its output.  Its cost is \$1595 on the Mini-Circuits website, however, we purchased this unit secondhand from an online auction website for approximately half this price.  Note that \mbox{LZY-22+} amplifier is designed to be powered using a \SI{24}{\volt}/\SI{5.5}{\ampere} DC power supply.  The VOLTEQ HY3006D \SI{30}{\volt}, \SI{6}{\ampere} DC power supply, for example, can be purchased for \$120 \mbox{(\url{http://www.volteq.com/})}.

{\it Bi-directional Coupler} -- Our experiments used the OSR Broadcast Research C21A8 bi-directional coupler.  Its useful frequency range is \SI{15}{\kilo\hertz} to \SI{1500}{\mega\hertz}, has a power rating of \SI{200}{\watt}, and a coupling of \SI{40}{\decibel}. OSR Broadcast Research no longer has an online presence.  A substitute bi-directional coupler from Mini-Circuits is the \mbox{ZABDC50-150HP+} which costs \$90.  Its useful frequency range is \SIrange[range-phrase=--, range-units=single]{0.4}{15}{\mega\hertz}, has a power rating of \SI{100}{\watt}, and a coupling of \SI{50}{\decibel}.

{\it Power Meter/Sensor} -- Our power measurements were made using Boonton 42B power meters with Boonton 41-4E power sensors.  The Mini-Circuits \mbox{ZX47-50-S+} power detector can be used from \SI{10}{\mega\hertz} to \SI{8}{\giga\hertz} to measure powers from $-45$ to $\SI[retain-explicit-plus]{+15}{dBm}$ and costs \$90.  Most manufacturers quote a minimum usable frequency of \SI{10}{\mega\hertz} for their power sensors.  Our measurements typically spanned \SIrange[range-phrase=--, range-units=single]{2.5}{5.5}{\mega\hertz} and we have found that reliable power measurements can still be made using these sensors.  Digital oscilloscopes with bandwidths above \SI{10}{\mega\hertz} are now common.  An oscilloscope with a voltage resolution of \SI{1}{\milli\volt} could be used to measure powers across a \SI[number-unit-product=\text{-}]{50}{\ohm} load as low as \SI{20}{\nano\watt} (\SI{-47}{dBm}).

{\it \SI[number-unit-product=\text{-}]{50}{\ohm} Termination} -- For all high-power measurements, the receiving loop was terminated using either a \SI[number-unit-product=\text{-}]{60}{\watt} incandescent light bulb or a \SI[number-unit-product=\text{-}]{50}{\ohm} load.  The C3N50 N termination from Centric RF is a suitable \SI[number-unit-product=\text{-}]{50}{\watt} load that costs \$125 \mbox{(\url{http://centricrf.com/})}.

{\it Copper Tubing} -- Flexible \SI[number-unit-product=\text{-}]{1/4}{inch} cooper tubing used to form the transmitting and receiving coils can be found at most hardware stores.  The cost is approximately \$1 per foot.  The \SI[number-unit-product=\text{-}]{55}{gallon} industrial plastic drums that the copper tubing was wrapped around can also be found at many hardware stores and costs approximately \$70. 

{\it Miscellaneous} -- The WPT experiments also required \SI[number-unit-product=\text{-}]{18}{AWG} magnet wire to form the transmitting and receiving loops, capacitors to set the resonant frequency of the loops, coaxial cables, and various RF connector adapters.  All of these components can be purchased from Digi-Key Electronics \mbox{(\url{https://www.digikey.com/})}.

\begin{acknowledgments}

We gratefully acknowledge insightful discussions with Thomas Johnson.

\end{acknowledgments}

\end{document}